\documentclass[%
aip,%
jcp,%
amsmath,amssymb,%
preprint,%
]{revtex4-1}

\usepackage{graphicx}
\usepackage{dcolumn}
\usepackage{bm}
\usepackage{color}

\begin{document}

\title{Single-file mobility of water-like fluid in a generalized Frenkel-Kontorova model}
\author{Patricia Ternes}%
\email{patricia.terdal@gmail.com}%
\affiliation{Instituto de F\'isica, Universidade Federal do Rio Grande do Sul, 
Caixa Postal 15051, 91501-970, Porto Alegre, RS, Brazil}%
\author{Alejandro Mendoza-Coto}
\email{alejandro.mendoza@ufsc.br}
\affiliation{Departamento de F\'isica, Universidade Federal de Santa Catarina, 88040-900 Florian\'opolis,
Santa Catarina, Brazil} 
\author{Evy Salcedo}%
\email{evy.salcedo.torres@ufsc.br}
\affiliation{Coordenadoria Especial de F\'isica, Qu\'imica e Matem\'atica, Universidade Federal de Santa 
Catarina, Rua Pedro Jo\~ao Pereira, 150, 88905-120, Ararangu\'a, SC,  Brazil}%

\date{\today}%

\begin{abstract}
In this work we used a generalized Frenkel-Kontorova model to study the mobility of water molecules 
inside carbon nanotubes with small radius  at low temperatures. Our simulations show that 
the mobility of the confined water decreases monotonically increasing the amplitude of the substrate 
potential at fixed commensurations. On the other hand, the mobility of the water molecules shows a 
non-monotonic behavior when varying the commensuration. This result indicates that the mobility of the 
confined fluid presents different behavior regimes depending on the amplitude of the water-nanotube 
interaction. In order to understand qualitatively these results, we study analytically the driven 
Frenkel-Kontorova model at finite temperatures. This analysis allows us to obtain the curves of the 
mobility versus commensurations, at fixed substrate potentials. Such curves shows the existence of three 
regimes of mobility behavior as a function of the commensuration ratio. Additionally, our study indicates 
a nontrivial and strong dependence of the mobility with a quantity that can be interpreted as an 
effective amplitude of the substrate potential, depending on the bare amplitude of the substrate 
potential, the commensuration ratio and temperature.
\end{abstract}

\maketitle
\section{Introduction}
A fluid in nanoscale confinement can present a dramatic change in its properties when compared with its 
bulk behavior. This occurs mainly due to the confining surface that, in this scale, is structured and 
smooth.

Single-wall carbon nanotubes (CNTs) have a periodic structure~\cite{iijima2} which allows the confined 
fluid to present anomalous behavior.~\cite{preenchimento1, preenchimento2} Moreover, they present
excellent electronic and mechanical properties, which makes them a promising nanochannel for a variety of 
applications in nanotechnology, specially in the biotechnology field.~\cite{remedio1,remedio2,remedio3,
vacina,cancer,implante} Therefore, the behavior of water confined inside CNTs has been extensively 
studied. This studies showed that, under confinement, water presents an anomalous flow when compared with
predictions of classical hydrodynamic theory.~\cite{fluxoExperimental1,fluxoExperimental2,experimentoBom,
fluxo1,teorico1,teorico2,teorico3,KOH17}

However, this anomalous behavior is not entirely understood. Experimental results showed that the water 
flow enhancement factor can vary from one to three orders of magnitude,~\cite{fluxoExperimental1,
fluxoExperimental2,experimentoBom} while simulations studies shows an enhancement of up to five orders of 
magnitude.~\cite{fluxo1,teorico1, teorico2,teorico3} Beyond the results range issue, there is not 
consensus about the enhancement factor behavior. Some previous works present a monotonic enhancement with 
decreasing the radius,~\cite{fluxoExperimental1, teorico2} while others present a discontinuous 
enhancement factor.~\cite{experimentoBom, teorico3} A simple explanation to the water mobility behavior 
was proposed by Falk et al.~\cite{atrito} They found that the nanotube curvature, given by the inverse of 
the nanotube radius, has a strong effect on water-CNT commensuration ratio: for more curved nanotubes she 
found more misfits water molecules over internal nanotube surface, leading to a higher mobility; while 
for a flat surface (in slab geometry) the mobility is independent of confinement (slab separation). So, 
for Falk et al., the high mobility is mainly related with the curvature-induced water-carbon 
incommensurability.~\cite{atrito}

The water has a complex phase diagram and there is no model to describe its behavior entirely. Several 
atomistic models were proposed in order to describe water's behavior, usually at ambient pressure and 
temperature.~\cite{h2o1, h2o2} Besides these models, effective models have been widely employed in order 
to study anomalous behavior in bulk water-like systems.~\cite{rafa16, rafa17,rafa18, rafa19, rafa24} In 
these effective models is considered that bulk water form small clusters of four water molecules, and 
that these clusters come together to form water bicyclo-octamers of different densities. However, in 
confined systems, these structures are completely different and heavily dependent of the tube geometry 
and water model employed.~\cite{preenchimento1, experimentoBom, estruturas, agua1,estruturas2} For CNTs 
with radius sufficiently small the confined water molecules form single-file chains, an arrangement that 
is an almost one-dimensional and highly oriented hydrogen-bonded network. Therefore, the choice of the 
inter-particle water potential is not trivial.

Understanding the water flow inside CNTs is a very complex task because the current available technology 
has a limited capacity to directly ``look'' at the dynamics of the confined molecules.~\cite{GUO15,BOC16}
So, to model the water-CNT system, theoreticians need to have some insight from a known nanoscopic system,
for example, the dynamics of atoms or molecules on a surface. For study the nanoscale friction several 
experimental methods have been developed to measure directly or indirectly the mobility of atoms or 
molecules on a surface. In this context, it is preferable to talk in terms of friction and mobility, 
rather than flow.

The most prominent apparatus developed to study the nanoscale friction was the atomic force microscope 
(AFM). AFM studies show that on the nanoscale the same physical system can exhibit a dry friction (like 
solid-solid friction) or wet friction (like liquid-solid friction), depending on several factors as 
module of shear stress, applied time of shear stress, etc.~\cite{MAT87,GEE90, GRA91} This behavior is 
completely different from that observed for macroscopic or even mesoscopic system. Through the quartz 
crystal microbalance (QCM), a novel experimental method developed by J. Krim, it was observed 
systematically that wet friction can indeed occur in structured systems.~\cite{ATRexp2,ATRexp1,ATRexp3} 
The results of QCM introduce a new key question to be answered, what is the main way of dissipation of 
energy in the nanoscale wet friction: the internal excitation due the interactions with the surface 
topology (phononic friction),~\cite{ph1,ph2,ph3,ph4} or the process of excitation of low-energy 
electron-hole pairs on the surface (electronic friction)?~\cite{el1,el2,el3,el4,el5,el6} To answer this 
question several and beautiful arranges were developed (from experimental perspective), even systems on 
superconducting state were analyzed.~\cite{ATRexp3,ALT12} From theoretical perspective almost all work 
was address by molecular dynamics simulation using a generalized Frenkel-Kotorova (FK) model as a 
paradigm.~\cite{ph1,ph2,ph3,Liebsch99,Tor1}

Our aim in this work is to study the relation between confined water mobility and the CNT topology. To 
do this, we performed molecular dynamics simulations using a generalized Frenkel-Kontorova model. The 
potentials were chosen to effectively reproduce the water behavior when confined in carbon nanotubes with 
very small radius (single-file state). The effect of the nanotube can be illustrated by a naive analogy 
with an egg carton, being the energy landscape smoother for nanotubes with smaller radius,~\cite{atrito} 
and the behavior of single-file water can be approximated as a one-dimensional chain of particles 
interacting through a two length scale isotropic potential.
To capture the topological effects on water mobility using the FK model, simulations should 
be performed in the low temperature regime. To avoid the water freezing, perfect commensuration ratios 
(0.5, 1.0) were avoided as well. Additionally, using an analytical analysis we found a relation between 
the mobility's behavior as a function of temperature (as predicted phenomenologically~\cite{BRA94,BRA96}) 
and corrugation. This paper is organized as follow: in section \ref{cap_fk_model} we present the 
generalized Frenkel-Kontorova model used in the characterization of the mobility of a confined water-like 
fluid. In section \ref{cap_results} we present and discuss the simulational results. To continue, section 
\ref{cap_teoria} shows an analytical analysis of the FK model at finite temperatures. Finally, in section 
\ref{cap_concl} our conclusions are presented.

\section{A Generalized Frenkel-Kontorova model}\label{cap_fk_model}

In this section we present the model used to study the dynamics of water molecules in a single file 
structure flowing inside a carbon nanotube. Our system of particles is study by means of a one-dimensional 
generalized Frenkel-Kontorova model. In this way the single file structure is regarded as a chain of 
particles with mass $m$, interacting through a core-softened potential $V\left( x_{ij}\right)$ (Eq. 
\ref{eqV}), where $x_{ij}=x_i-x_j$, represents the distance between particles $i$ and $j$. Moreover we 
modeled the influence of the structured nanotube by the presence of an external periodic potential, 
$U\left(x_{i}\right)$ (Eq. \ref{eq_pot_sub}). To induce a net flow of particles in a given direction we 
applied an external force $F$ on each particle of the system. 

Additionally, the temperature is fixed adding a stochastic force, $f_i(t)$, and a viscous damping force, 
$-m\eta_{ad}\dot{x_i}$. The stochastic force is related with temperature and with the 
\textit{ad hoc} damping constant ($\eta_{ad}$) by means of the fluctuation-dissipation 
theorem: $\left\langle f_{i}\left(t\right)f_{j}\left(0\right)\right\rangle =2\eta_{ad}mk_{B}
T\delta_{ij} \delta\left(t\right)$. Here $k_{B}$ is the Boltzmann constant and $\eta_{ad}$ 
can be thought as resulting of the influence of certain degrees of freedom inherent of real physical 
systems that are not included in our model. With such ingredients the equation of motion of our system is 
given by the following  Langevin equation:
\begin{equation}
m \ddot{x}_{i}(t) + \eta_{ad} \dot{x}_i(t)= -\frac{dU\left(
x_{i}\right) }{dx_{i}} -  \frac{dV\left( x_{ij}\right) }{dx_{i}}  + f_{i}
\left(t\right) + F_i.
\label{eq_modelo}
\end{equation}

The interaction between particles, $V\left( x_{ij}\right)$, is a potential obtained by the addition of 
$3$ different Fermi-Dirac distributions:\cite{evy}
\begin{equation}
V\left( x_{ij}\right)  = \sum_{k=1}^3 \frac{\varepsilon_k}{\exp \left(\frac{x_{ij} - 
r_{0k}}{\sigma _k} \right)  + \alpha_k},
\label{eqV}
\end{equation}
where $x_{ij} = |x_j - x_i|$ is the distance between two water-like particles. Table \ref{tab_FERMI} shows 
the parameters used for $V\left( x_{ij}\right)$ in our simulations. The periodic potential $U\left( x_{i}
\right)$ representing the water-CNT interaction is:
\begin{equation}
U\left(x_{i}\right)=u_{0}\cos\left(\frac{2\pi}{a}x_{i}\right),
\label{eq_pot_sub}
\end{equation}
being $a$ and $u_{0}$ the periodicity and the amplitude of the potential respectively. The amplitude of 
potential is also called corrugation. We choose the parameters for the potentials $V\left( x_{ij}\right)$ 
and $U\left( x_{i}\right)$  to effectively reproduce the water-water and water-CNT interactions for the 
TIP3P water model confined in a nanotube. The complete discussion for these potentials determinations can 
be found in the supplemental material at section \ref{supplementary}.

In what follows we take the equilibrium distance between two oxygens of molecules forming a hydrogen bond 
as $r_{I}$, the energy of the hydrogen bond as $E_{I}$, and the mass $m$ of a water molecule as the 
fundamental units of the problem. Consequently all physical quantities are expressed in reduced units, 
this is:
\[
\begin{matrix}
  x^{*} &\equiv& x r_{I}^{-1}, 	  	&\hspace{50pt}\ & F^{*} &\equiv& F r_{I}E_{I}^{-1}, \\
  T^{*} &\equiv& T k_B E_{I}^{-1},	&\hspace{50pt}\ & t_{0} &\equiv& r_{I}(m / E_{I})^{1/2}, \\
   V^{*}&\equiv& V E_{I}^{-1},		&\hspace{50pt}\ & t^{*} &\equiv& t t_{0}^{-1},
\end{matrix}
\]
where $t$ is time.
\begin{table}[h!]
\begin{center}
  \caption{Parameters for the potential $V(x_{ij})$ in reduced units of $m$, $E_{I}$ and $r_{I}$.}
  \begin{tabular}{ l  l l l }
    \hline \hline
    $\varepsilon_{1,2,3}^{*}$	& 1.0		& -0.001	& -0.999
    \\ $r_{0_{1,2,3}}^{*}$	& 0.940		& 1.025		& 1.105
    \\ $\sigma _{1,2,3}^{*}$	& 0.005		& 0.005		& 0.005
    \\ $\alpha_{1,2,3}$		& 0.0		& 1.0		& 1.0
    \\ \hline \hline
  \end{tabular}
\label{tab_FERMI}
\end{center}
\end{table}

The length of the chain of particles is fixed, consequently in order to impose periodic boundary 
conditions we have the following constraint: $Ma=Nr_{I}$, where $a$ was defined in Eq. \ref{eq_pot_sub},
$N$ corresponds to the number of particles in our chain and $M$ represent the number of minimums of the 
potential within the chain. Now we define the so called commensuration ratio:
\begin{equation}
 \zeta= \frac{a}{r_I}=\frac{N}{M}.
\end{equation}

Since $E_{I}$ and $r_{I}$ were taken as fundamental units, changes in the structure of the confined water
are reflected in a commensuration variation while little changes in the CNT geometry are treated through
a corrugation variation.~\cite{atrito} Therefore, $u_0$ and $\zeta$ are our key parameters in the 
mobility study of the chain particles.

\begin{figure}[h!]
\begin{center}
  \includegraphics[width=2.5in]{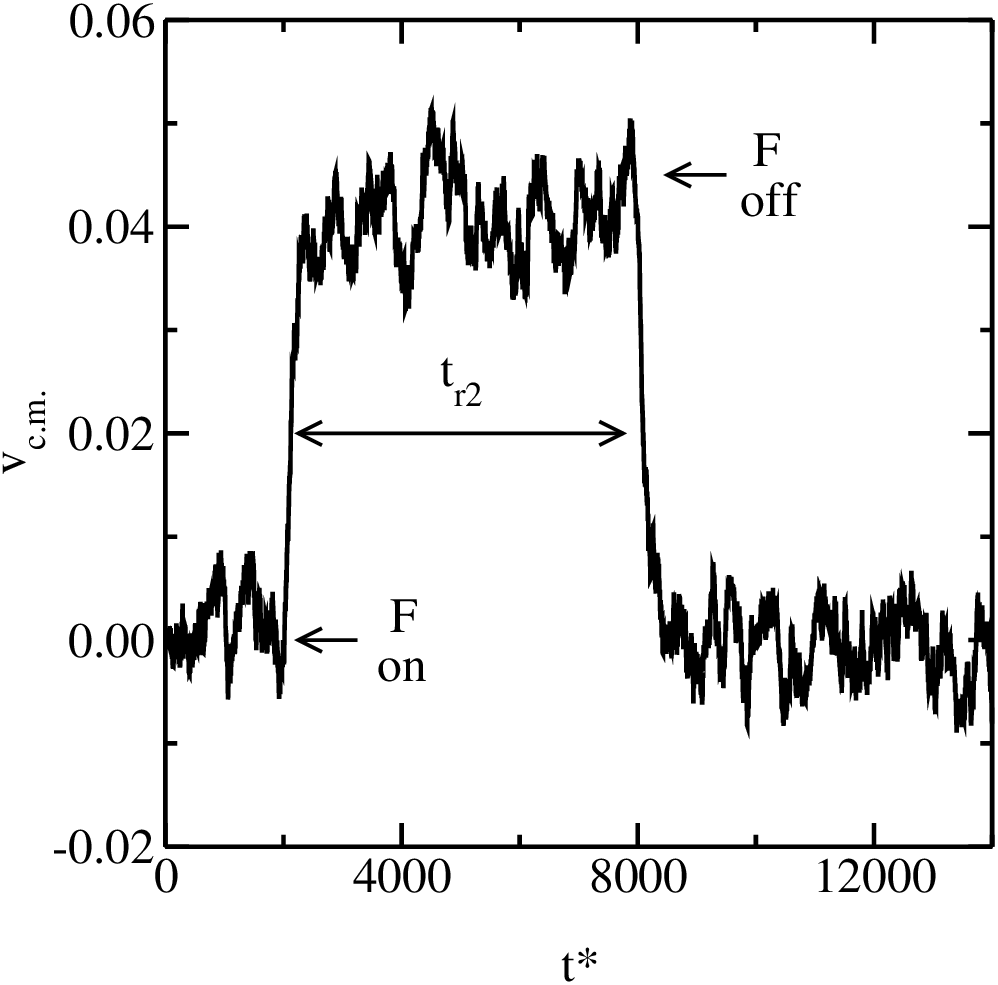}
  \caption{Typical behavior of the center-of-mass velocity of the chain particles versus time. In this 
	   case $\zeta = 0.68$ and $u_0^{*}=0.04$.}
  \label{fig_typicalVCM}
\end{center}
\end{figure}

Equation \ref{eq_modelo} is numerically integrated using a brownian molecular dynamics algorithm,~\cite{
Allen} for $N = 4096$ particles with a time step $dt =0.001t_0$ at $T^{*} = 0.05$. We allow the system to 
relax during a time interval of $2000t_0$, in the sequence an external force $F^{*}=0.0003$ is applied; 
after another transient period of order $6000t_0$ the system reaches the steady state and the external 
force is turned off making the velocity of particles decay back to zero. Figure \ref{fig_typicalVCM} 
shows the typical behavior of the center-of-mass velocity of the chained particles during simulation.

The autocorrelation function of the velocities between several different samples, in the decay region 
(after the driving force is turned off), has the following exponential form:~\cite{correlation}
\begin{equation}
\gamma(t-t') = \exp\left(-\eta |t-t'|\right),
\label{eq_expDecay}
\end{equation}
where $\eta$ will be the effective microscopic friction coefficient. An exponential decay of 
the velocity autocorrelation function as the one given by equation \ref{eq_expDecay} is characteristic of 
fluid on solid systems.~\cite{BHU95} We expect that the measured $\eta$ to be higher than the 
ad hoc damping constant ($\eta_{ad}$), which was set it to $5\times 10^{-3}t_0^{-1}$. Finally all 
simulations were carried out at fixed temperature, volume and number of particles.

\section{Computational Results}\label{cap_results}
We carried out simulations with several commensuration ratios between $0.55$ and $0.91$ and corrugation 
values from $0.01$ to $0.10$. For all these parameters, a wet friction was observed. In all 
cases the coefficient of friction was obtained from fits according to Equation \ref{eq_expDecay}. To 
calculate the velocity autocorrelation function we used fifty different samples of the velocity 
relaxation.

\begin{figure}[h!]
\begin{center}
  \includegraphics[width=2.5in]{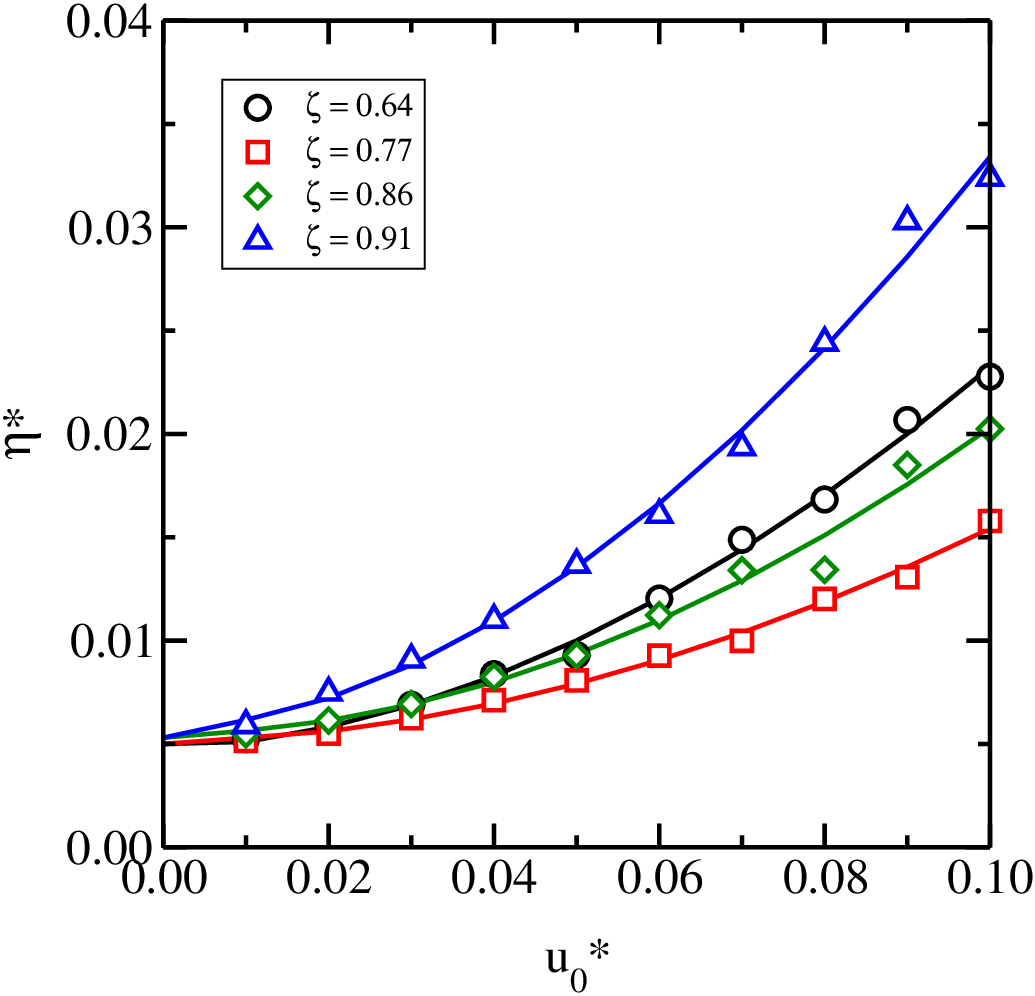}
  \caption{Total friction coefficient for some commensurations $\zeta$: black circle
	  $0.64$, red square $0.77$, green diamond $0.86$ and blue triangle up 
	  $0.91$. The solid lines are fits to Equation \ref{eq_powerLaw}.
	  Error bars are omitted, because they are smaller than the size of the symbols.}
  \label{fig_uXeta}
\end{center}
\end{figure}

Figure \ref{fig_uXeta} shows the total friction coefficient as a function of the corrugation for some 
commensuration ratios. These results show that as the corrugation increases the friction experienced by 
the water-like particles increases. An analysis of the curves $\eta \times u_0$ shows that these curves 
presents a power law behavior given by:
\begin{equation}
\eta = \eta_{0}+cu_0^n,
\label{eq_powerLaw}
\end{equation}
furthermore in the limit of corrugation going to zero $\left(u_{0}\rightarrow0\right)$ the coefficient 
$\eta_{0}$ has an average value equal to the \textit{ad hoc} damping constant: $\eta_{0}=
0.005\pm 0.001$. Therefore, it is natural to assumed that the topological influence in the 
mobility is represented by the second term of equation \ref{eq_powerLaw}: $cu_0^n$, where the average 
value of $n$ obtained from the best fit is equal to $1.9 \pm 0.3$. 
It is worth noticing that a number of previous analytical and numerical works report a quadratic behavior
of the friction with the corrugation amplitude: for fluid on surface,~\cite{ph1, ph2,Tor1,Liebsch99} 
for a fluid slab confined between parallel solid walls,~\cite{BAR99} and for TIP3P water confined in 
carbon nanotubes.~\cite{atrito}

\begin{figure}[h!]
\begin{center}
\includegraphics[width=3.37in]{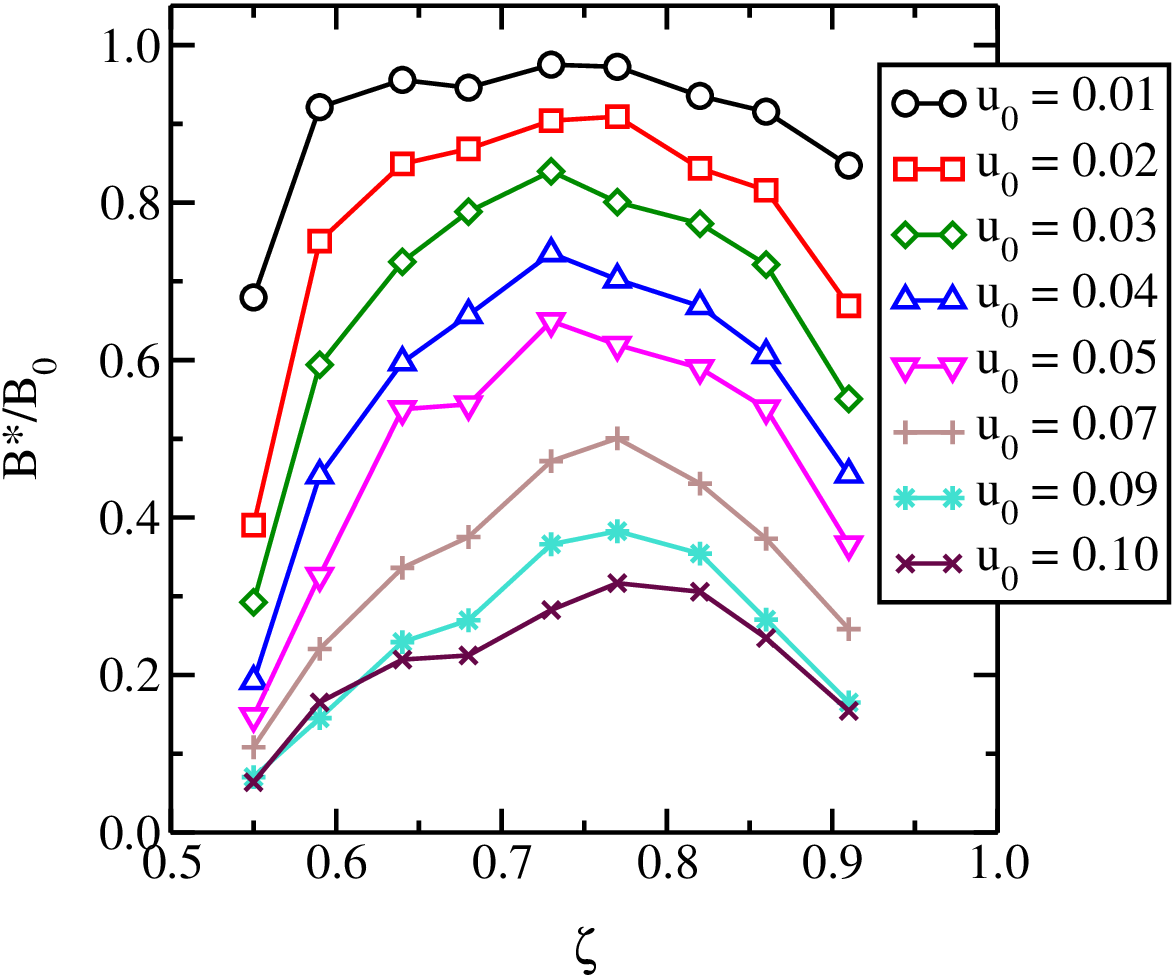}
  \caption{Normalized mobility in function of commensuration ratio for some substrate potential amplitude 
  $u_0^{*}$, where $B_0$ is the mobility in the limit of absence of corrugation $\left(u_{0}\rightarrow0
  \right)$. Error bars are omitted, because they are smaller than the size of the symbols.} 
  \label{fig_bXzeta}
\end{center}
\end{figure}

The central quantity in our study is the mobility, $B$, of the particles chain, which is 
defined as the inverse of the friction coefficient $\eta$: $B=\eta^{-1}$.
In Figure \ref{fig_bXzeta} we showed the normalized mobility as a function of the commensuration ratios.
The mobility decreases as the corrugation increases for all studied corrugation amplitudes, but the 
relation between commensuration and mobility is non-monotonic. The mobility as expected does have local 
minimum  for commensuration rations close to $1/2$ and $1$. For commensuration ratios between these 
values the mobility increases reaching a local maximum value in a region close to $\zeta = 3/4$. In 
addition, for corrugations higher than $u_0=0.05$, the mobility presents a change of curvature at $\zeta$ 
close to $2/3$. These results indicate the existence of different mobility regimes depending 
on the amplitude of interaction between water-like particles and CNT surface.

At this point it is interesting take into consideration the work of Braun and Kivshar,~\cite{BRA94} which
study the behavior of the mobility as a function of temperature and commensurations, at fixed corrugation 
for the Frenkel-Kontorova model. Although our results for the mobility were obtained for a different model 
and varying corrugation and commensuration at fixed low temperature, we will show shortly that there is a 
close relation between our results and those of Braun and Kivshar.~\cite{BRA94}

In that work the authors present a  phenomenological approach that describe the transport properties in 
terms of moving defects (kinks) in the density of particles along the chain. Kinks describe local 
expansions or compressions of the chain which are at the core of the mechanism producing the mass 
transportation. It is expected that the higher the concentration of kinks the greater the mobility. 
Accordingly to reference~\cite{BRA94} at high temperatures the concentration of kinks produced due to 
thermal fluctuations has a maximum at $\zeta=3/4$, consequently the mobility have a local maximum in this 
region.

For lower temperatures the concentration of kinks depends on the degree of frustration in the system's 
geometry. When the mean distance between particles coincides with the distance between two minima of the 
external periodic potential ($\zeta = 1/2$ and $1$) the system is in a commensurate ground state and 
particles are pined. For commensurations between these values there is a nonzero probability to find 
particles close to the maximum of the external potential which increases the kinks concentration. This 
produces an increment of the particles mobility. For commensuration equal to the golden mean $\zeta=2/
(1+\sqrt(5))$ the geometric frustration is maximal as well as the kinks concentration, therefore a local 
maximum in the mobility occurs. However, in this conditions, for sufficiently low temperature due to the 
high density of kinks a regular lattice of kinks is formed decreasing the mobility strongly. This behavior 
is observed first for commensurations close to the Golden mean. From this state a further decrease of 
temperature produce similar falls in mobility for commensurations such as $3/4$, $3/5$ or $4/5$.~\cite{
BRA96}

As we mention before, this phenomenological description is directly related to changes in temperature. However
our computational results were obtained keeping temperature fixed at a low value while 
the corrugation was varied. Even so, some similarities arise. The Figure \ref{fig_bXzeta} 
presents two important results. First, the mobility depends on the commensuration ratio  as described by
Braun and Kivshar. .~\cite{BRA94,BRA96} Second, the mobility  presents different regimes depending on the substrate 
amplitude. From the theoretical background at nanoscale friction, we have different mobility regimes due to 
the activation of geometric or thermal kinks, depending on the system temperature The similarities between 
the Braun and Kivshar results for the mobility at constant corrugation varying temperature,~\cite{BRA94} 
and ours obtained varying corrugation at fixed temperatures motivates us to pursue further investigation of 
the corrugation versus temperature interplay. The main objective of this investigation is to 
Provide an explanation for the existence of different mobility regimes as a function of the corrugation observed 
in Figure \ref{fig_bXzeta}.

\section{Theoretical Discussion}\label{cap_teoria}
In order to explain the previously obtained results for the mobility, in both the classic and in our 
generalized Frenkel-Kontorova model, we decided to study analytically the behavior of the mobility in the 
classic FK model at finite temperatures. We begin considering the following equation of motion for the 
system of interacting particles:
\begin{equation}
 m\ddot{x_i}+\eta\dot{x_i}=K(x_{i-1}+x_{i+1}-2x_i)-\frac{dU}{dx}(x_i)+F+f_i(t),
\end{equation}
where $U(x)$ represent the substrate potential taken as $U(x)=-u_0\cos(k_0x)$. Choosing now $l_0=k_0^{-1}$ 
and $\sqrt{m/K}$, as units of length and time respectively, we can rewrite our system of equations in 
terms only of dimensionless variables:
\begin{equation}
\ddot{r_i}+\tilde{\eta}\dot{r_i}=(r_{i-1}+r_{i+1}-2r_i)-g\sin(r_i)+\tilde{F}+\tilde{f}_i(\tilde{t}),
\label{dless}
\end{equation}
where $r$ and $\tilde{t}$ represent the dimensionless position and time, while $\tilde{\eta}=\eta/
\sqrt{mK}$, $g=u_0k_0^2/K$, $\tilde{F}=k_0F/K$ and $\tilde{f}_i(\tilde{t})=k_0f_i(t)/K$. In what follows 
we will omit tildes in dimensionless quantities to simplify the notation. 

In order to calculate analytically the stationary mobility of the system, defined as the ratio of the 
driving force divided by the average stationary velocity of the particles, we need to determine the 
stationary state of the system described by Eq.\ref{dless}. A procedure to find such stationary state have 
been developed in reference~\cite{StEl98}. Nevertheless such technique can not be applied directly in our 
case due to the presence of the thermal noise. However, if we perform an average over thermal noise 
realizations in each term of equation \ref{dless}, we will recover a deterministic equation for which 
techniques developed by Strunz and Elmer~\cite{StEl98} can be applied. Performing such average we reach to: 
\begin{eqnarray}
\nonumber
 \ddot{\langle r_i\rangle}+\eta\dot{\langle r_i\rangle}&=&(\langle r_{i-1}\rangle+\langle r_{i+1}
 \rangle-2\langle r_{i}\rangle)\\ &-&g\langle\sin(r_i)\rangle+F.
\label{teq}
\end{eqnarray}
To proceed we need to calculate the thermal average of $\sin(r_i)$. This term can be evaluated by using 
the following mean field argument: first we consider that the stationary sliding state is uniform and in 
this sense this means that within the mean field approximation, we could use a single particle 
Fokker-Planck equation to study the dynamics of a single particle density and conclude that within such 
approximation the average time dependent single particle probability distribution will be:
\begin{equation}
P(x,t)=\sqrt{\frac{K}{\pi k_BT}}\exp\left(-\frac{k(x-vt)^2}{k_BT}\right).
\label{prob}
\end{equation}
To reach such a result it is implicit that in thermal equilibrium the average distance between particles
is equal to the equilibrium distance between them in the absence of all external forces. With equation 
\ref{prob} in mind we can conclude that:
\begin{equation}
\langle \sin(r_i(t))\rangle=\sin(\langle r_i(t)\rangle)\exp\left(-\frac{k_0^2k_BT}{4K}\right).
\end{equation}
This relation implies that Eq. \ref{teq} can be written as:
\begin{eqnarray}
\nonumber
 \ddot{\langle r_i\rangle}+\eta\dot{\langle r_i\rangle}&=&(\langle r_{i-1}\rangle+\langle r_{i+1}\rangle-
 2\langle r_{i}\rangle)\\  &-&g_r\sin(\langle r_i\rangle)+F,
\label{dteq}
\end{eqnarray}
where $g_r=g\exp\left(-\frac{k_0^2k_BT}{4K}\right)$. Interestingly these results implies that the dynamics
of the driven Frenkel-Kontorova model at finite temperatures is equivalent to the one at zero temperature, 
with a modified effective potential substrate amplitude. Now we can apply directly the techniques developed 
by Strunz and Elmer~\cite{StEl98} to study the stationary uniform sliding state of equation \ref{dteq}. 

The afore mentioned method propose an anzats for the stationary form of the position of the particles as
a function of time. The proposed functionality is given by:
\begin{equation}
 r_i=\psi+ai+vt+f(\psi+ai+vt),
\end{equation}
where $\psi$ represent an arbitrary phase, $a=k_0l$ is the equilibrium distance between particles in 
units of $k_0^{-1}$ and $v$ corresponds to the stationary velocity of the system of particles. Moreover 
the function $f(x)$ is the so-called hull function, which is a zero mean periodic function characterizing
the oscillation of the particle over its uniform translational movement due to presence of the periodic 
potential. 

To continue we substitute the previous anzats for $r_i(t)$ into our equation of motion (Eq. \ref{dteq}). 
This procedure leads us to the following differential equation for the hull function.
\begin{eqnarray}
\nonumber
 v^2f''(\phi)+\eta v(1+f'(\phi))&=&(f(\phi-a)+f(\phi+a)-2f(\phi))\\
 &-&g\sin(\phi+f(\phi))+F
 \label{eqf}
\end{eqnarray}
Considering now that $f(\phi)$ must be a periodic function with zero mean we know that we can write the 
hull function as: $f(\phi)=\sum_{n=1}^\infty a_n\cos(n\phi)+b_n\sin(n\phi)$, where $a_n$ and $b_n$ 
represent the Fourier expansion coefficients of the hull function. If the Fourier expansion for 
$f(\phi)$ is substitute in equation \ref{eqf} we can reach to the following system of equation for the 
Fourier coefficients:
\begin{eqnarray}
\nonumber
 &&\left[v^2n^2+2(\cos(na)-1)\right]a_n-\eta vnb_n=\\ \nonumber
 &&g\int_0^{2\pi}\frac{d\phi}{\pi}\sin(\phi+f(\phi))\cos(n\phi)\\  \nonumber
 &&\left[v^2n^2+2(\cos(na)-1)\right]b_n+\eta vnb_n=\\  
 &&g\int_0^{2\pi}\frac{d\phi}{\pi}\sin(\phi+f(\phi))\sin(n\phi). 
 \label{seq}
\end{eqnarray}
Once the Fourier coefficients have been determined, the driving force $F$ can be calculated in terms of 
$v$ and the set of Fourier coefficients ${a_n,b_n}$. This relation can be proven to be:~\cite{StEl98}
\begin{equation}
 F=\eta v\left(1+\frac{1}{2}\sum_{n=1}^{\infty}n^2(a_n^2+b_n^2) \right).
 \label{fviz}
\end{equation}

Considering that the method outlined allows uets to calculate the values of stationary velocity $v$ for a 
given driving force $F$, now we can calculate the mobility of the system. The computational study 
presented focused on the role of the commensuration in the behavior of the mobility, at different 
amplitudes of the substrate potential. Now we will determine similar curves numerically for the 
Frenkel-Kontorova model for comparison with the computational results from simulations. 

To continue we need to write all our dimensionless parameters in terms of the commensuration parameter 
$\zeta$ defined as $2\pi/a$ and an appropriate energy scale, such energy scale is chosen as $Kl^2$. This 
lead us to the following relations:
\begin{eqnarray}
\nonumber
 g&=&u_0\frac{4\pi^2}{\zeta^2}\exp\left(-\frac{\pi^2}{\zeta^2}\tilde{T}\right)\\ \nonumber
 F&=&\frac{2\pi}{\zeta}\tilde{F} \\
 a&=&\frac{2\pi}{\zeta}
 \label{rel}
\end{eqnarray}
where $u_0=u/(Kl^2)$, $\tilde{T}=(k_BT)/(Kl^2)$ and $\tilde{F}=F/(Kl)$. To conclude we set $\eta=1$ and 
proceed with the numerical study.

\begin{figure}[h!]
\begin{center}
  \includegraphics[width=2.5in]{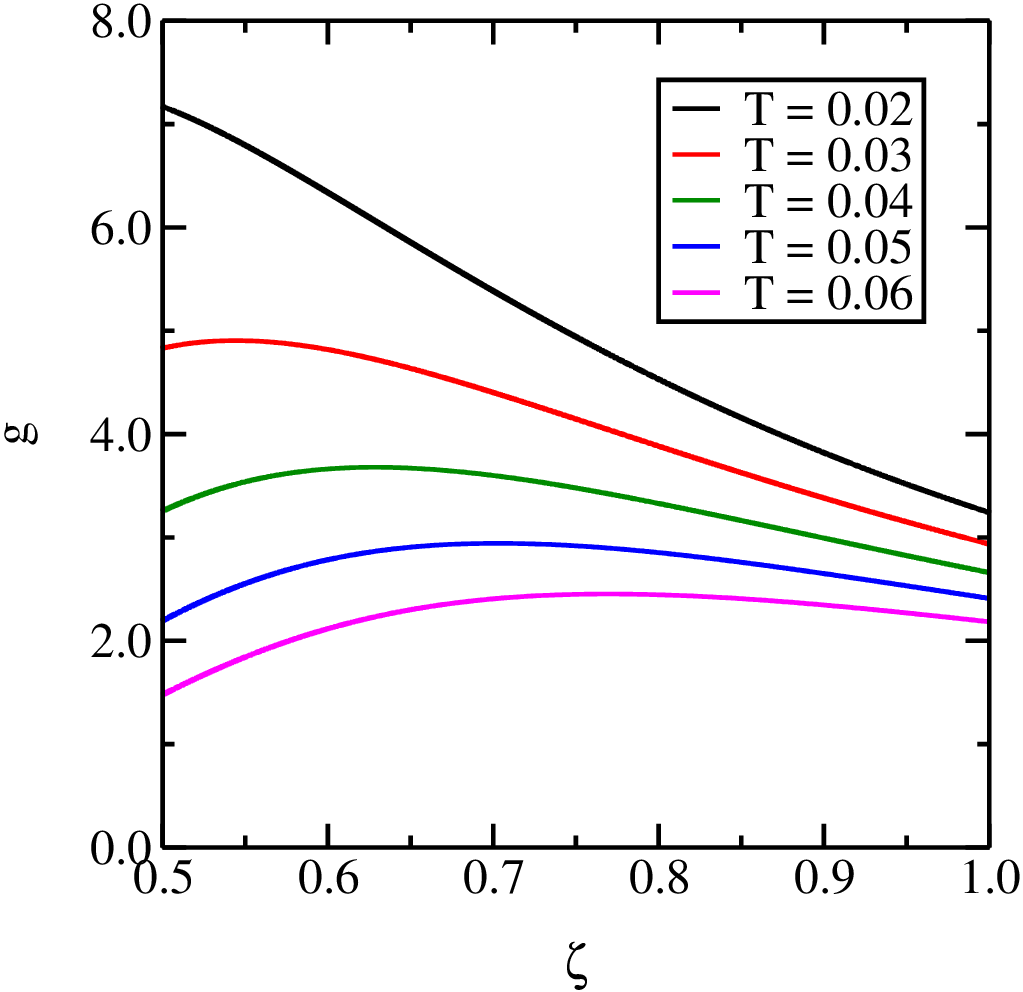}
  \caption{Effective substrate amplitude potential as a function of the commensuration at different 
           temperatures for $u_0=0.1$. As shown in the figure the effective substrate potential as a 
           function of $xi$ depends critically on temperature.}
  \label{potvsxi}
\end{center}
\end{figure}

The first thing that is worth noticing is the non-trivial dependence of the effective lattice potential $g$
in terms of the commensuration $\zeta$. As we can observe from Fig.\ref{potvsxi} the effective potential 
to which particles are submitted in general varies strongly with the commensuration and at the same time 
such variation depends critically on temperature. More important than this is to note from equations 
\ref{rel} that the effective corrugation $g$ depends on both the bare corrugation and the temperature. 
In fact at fixed commensuration the effective corrugation is an increasing monotonic function of $u_0$ and 
a decreasing monotonic function of temperature. Equations \ref{rel} clarifies the interplay between 
temperature and substrate potential and ultimately allows us to establish a connection between the 
behavior of the mobility, varying corrugation at fixed temperature and varying temperature at fixed 
corrugation. To explore the consequences of our analytical calculations we proceed with the calculus of 
the mobility curves varying the corrugation at fixed temperatures. 

Now we can describe the protocol followed for the numerical solution of equation \ref{seq}. As input 
parameters we have the friction coefficient $\eta$, set it to one, the geometrical parameter $a$, the 
effective corrugation potential $g$ and the stationary mean velocity $v$. With the numerical solution at 
hand the calculation of the driving force $F$ producing such stationary velocity can be carried out 
directly applying Eq.\ref{fviz}. This procedure is used to calculated the normalized mobility defined as 
$B=(\eta v)/F$ following Eq.\ref{fviz}. As can be noticed from this equation the normalized mobility is 
always a quantity lower than one, being one only in the limit of zero corrugation potential.  

\begin{figure}[h!]
\begin{center}
  \includegraphics[width=3.37in]{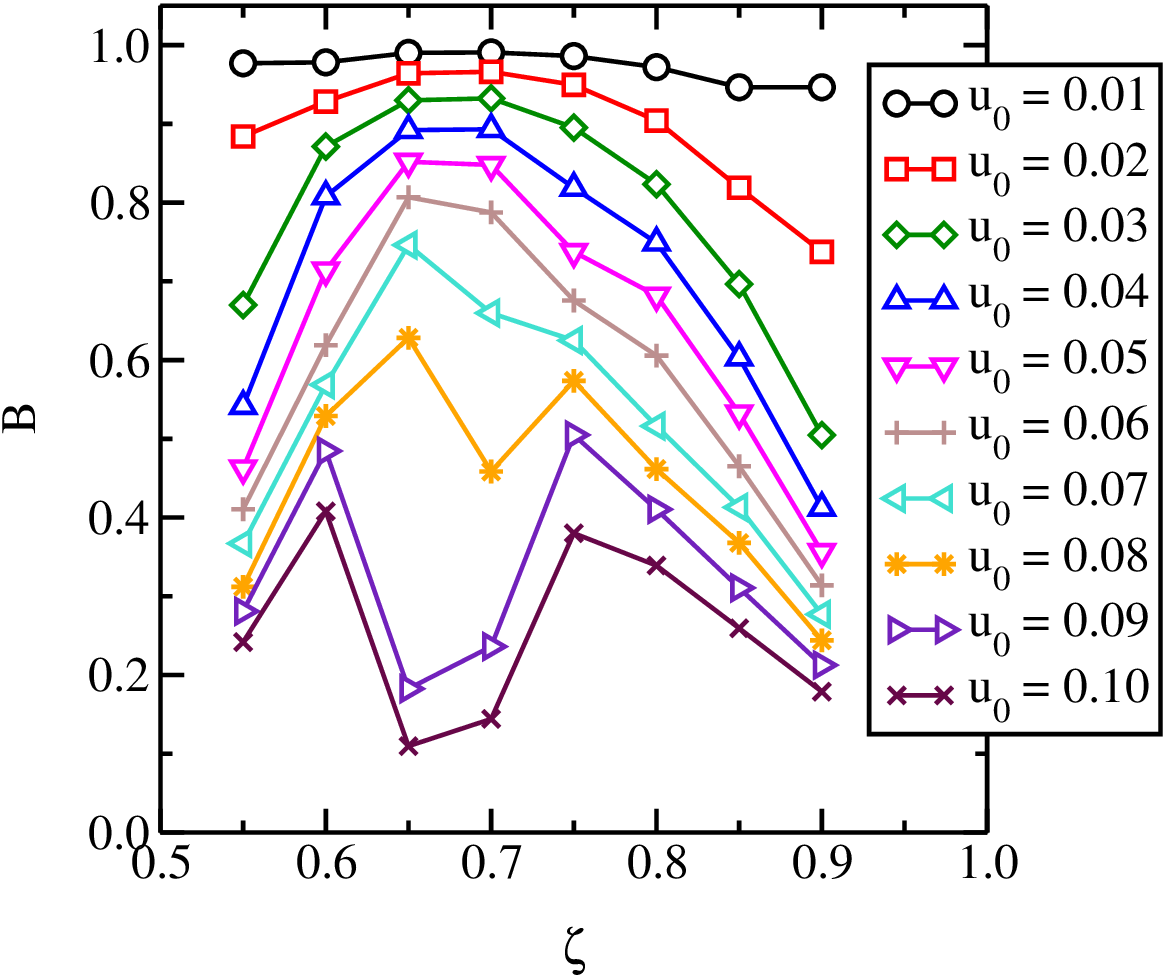}
  \caption{Normalized mobility at $\tilde{T}=0.038$ for different amplitudes of the substrate potential 
           $u_0$. All curves were calculated at a fixed driven force $\tilde{F}=0.13$.}
  \label{t2}
\end{center}
\end{figure}

\begin{figure}[h!]
\begin{center}
  \includegraphics[width=3.37in]{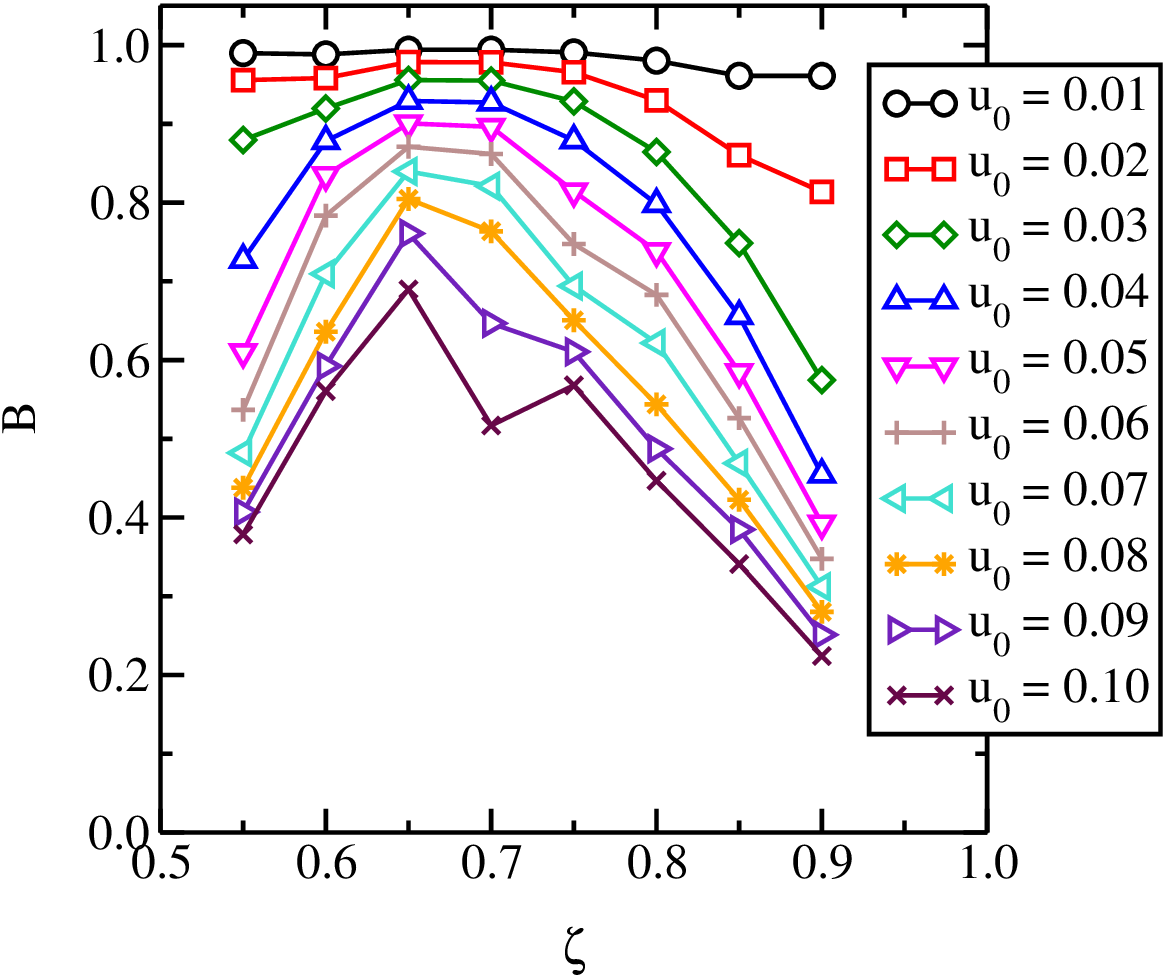}
  \caption{Normalized mobility at $\tilde{T}=0.05$ for different amplitudes of the substrate potential 
           $u_0$. All curves were calculated at a fixed driven force $\tilde{F}=0.13$.}
  \label{t1}
\end{center}
\end{figure}

We proceed next with the construction of families of normalized mobility curves varying commensuration, 
for various amplitudes of the corrugation potential at fixed temperature and driving force. We construct 
this families for two relatively low temperatures arbitrarily chosen: $\tilde{T}_1=0.038$ (Fig.\ref{t2})
and $\tilde{T}_2=0.05$ (Fig.\ref{t1}). At the same time the bare corrugation potential $u_0$ was let it 
range from $0.01$ to $0.1$ and the driving force in all cases was taken as $0.13$. It is worth noticing 
that although all numerical values were selected arbitrarily, since the values of the temperature 
strongly affects the effective value of the substrate potential, much higher temperatures will result in 
very small $g$ leading the system to a trivial regime out of the scope of the present work.

In figures \ref{t2} and \ref{t1} a number regularities appears, as a reflection of a robust physics over 
variations of temperature and substrate potential. To understand qualitatively the behavior of the 
mobility we make use of the picture given by Braun et al.~\cite{BRA94,BRA96} previously discussed in 
section \ref{cap_results}. First it is important to notice that the maximum of the mobility should be 
reached when the concentration of kinks is maximum as well. As we know there are two sources of kinks, 
those produced by the geometric frustration and those due to thermal fluctuations. According to Braun et 
al.~\cite{BRA94,BRA96} in the high temperature regime, or equivalently in the low substrate potential 
regime (as discussed previously), we have a maximum of the total number of kinks at commensurations 
around $\zeta=3/4$, reflecting in a maximum of the mobility at such commensurations, as observed in our 
figures \ref{t2} and \ref{t1}. 

Lowering temperatures, or increasing $u_0$, the population of thermally activated kinks decreases 
considerably making the geometrical frustration the leading mechanism to increase the kinks concentration.
As expected in this regime, a maximally geometric frustrated system should exhibit the maximum of the 
mobility. According to the literature the maximal geometric frustration occurs when the commensuration is 
equal to the golden mean, i.e. $\zeta=2/(\sqrt{5}+1)$.~\cite{Pey1} This fact explains why in figures 
\ref{t2} and \ref{t1} the mobility at intermediate values or substrate potential have a maximum at a 
value close to the golden mean. In such condition a further increase of the substrate potential, or 
analogously a decrease of temperature, produce a sort of solidification of the dense system of strongly 
interacting kinks, which reflects in a sudden decrease of the mobility exactly at the same 
commensurations where originally the mobility use to have a maximum. 

\section{Conclusions}\label{cap_concl}
In this work we used a generalized Frenkel-Kontorova model to study the mobility of water confined in a 
carbon nanotube with very small radius. In this highly confined regime the water molecules form an almost 
one-dimensional network of highly oriented particles linked by the hydrogen bound. We described such 
system as a one-dimensional chain of particles interacting through a core-softened potential of two 
length scales. 

We performed Langevin's simulations for a system of water-like particles and study the behavior of the 
total friction coefficient in low temperature regime. For all parameters adopted in this work,
we observed wet friction. Our computational results shows  a quadratic increasing monotonic behavior of 
the friction coefficient as a function of the corrugation, at fixed commensurations.
At the same time the friction coefficient shows a non-monotonic behavior varying commensurations at fixed 
amplitudes of the substrate potential. Such result for the friction coefficient is better observed in our 
family of mobility versus commensuration curves, for different corrugations. These two behaviors obtained 
corroborate the results reached through atomistic simulations of TIP3P water confined in carbon nanotubes
developed by Falk et al.,~\cite{atrito} where they concluded that the incommensurability between water 
and the carbon nanotube is the main result for the high flux observed in this system.
Additionally, although obtained from an effective model, our results opens a path for the understanding of 
the physical phenomenon responsible for the variety of results obtained for the water flow in CNT.~\cite{
fluxoExperimental1,fluxoExperimental2,experimentoBom,fluxo1,teorico1,teorico2,teorico3,KOH17,atrito} From 
a computational perspective, each atomistic model of water has its own geometric parameters. These small 
geometric  differences between the various water models result in different lengths for hydrogen bonds, 
for example. Therefore, different water models have different commensuration ratios with the nanotube, and 
as our results consistently show, even small changes in commensuration ratio, affects the mobility.

The mobilities curves not only shows a non-monotonic behavior of this quantity varying commensuration, but 
also indicates the existence of different regimes depending on the amplitude of the corrugated potential.
In order to understand qualitatively the computational results for the mobility we study analytically a 
simplified version of our model, the driven Frenkel-Kontorova model at finite temperatures. As a result 
we obtain numerically the curves of mobility versus commensurations at fixed substrate potentials which as 
discussed previously shows the existence of three regimes of behavior of such quantity with variations of 
commensurations. This regimes can be related with the density and dynamic of the defects in our chain of 
particles in a scenario analogous to the one proposed by Braun et al.,~\cite{BRA94,BRA96} considering a 
sort of inverse relation between temperature and amplitude of the substrate potential. 

In this sense, the main result of this work is the existence of a relation between the effective substrate
potential with temperature, commensuration and the bare substrate potential. This relation although 
simple shows that even in our simplified model there is a nontrivial and strong dependence of the 
effective substrate potential with the models parameters (temperature, commensuration and substrate 
potential), which at the same time makes the mobility of the system strongly dependent of this quantities.
This result adds one more ingredient to the discussion of the differences between works 
on water flow.~\cite{fluxoExperimental1,fluxoExperimental2,experimentoBom,fluxo1,teorico1,teorico2,
teorico3,KOH17,atrito} As we have already discussed, the commensuration ratio certainly plays an 
important role for the mobility, but others parameters of the system like temperature and corrugation are 
relevant as well.

When we compare the analytic and the simulational results for the mobility varying commensuration, at 
fixed substrate potential, some similarities arise. Although it is expected that the form of the 
potential exert a critical influence on the detailed form of the mobility curves, we can indeed identify 
in both cases a regime for small amplitudes of the potential in which we have a displacement of the 
maximum of the mobility from higher to lower values of the commensuration. This crossover was 
theoretically identify as a result of the variation in the relative population of geometric and thermal 
kinks. We expect the same mechanism to be responsible for the observed behavior in the generalized 
Frenkel-Kontorova model. Increasing the substrate potential our model for water-like particles develops a 
change in the curvature of the mobility curves evident as an additional drop in the mobility at 
commensurations close to the golden mean. This behavior in our generalized FK model could be explained by 
the same mechanism that produce a drop in the mobility for the FK model, i.e. a structuring of a large 
number of geometric induced defects in the system.  

Although we are aware that further studies are needed to confirm the proposed mechanisms as the 
responsible for the observed features of the mobility curves, in the generalize FK model, we expect that 
the present work motivates and shed some light on the physics involving the transport process of water 
particles in extreme confinement conditions.

Finally, we would like that this work motivates the use of the various experimental and 
theoretical methods developed originally for the study of sliding friction in nanoscale, to understand 
the water behavior in nano-confined conditions.
\section{Supplementary Material}\label{supplementary}
See the supplementary material for the complete description of 
$V\left( x_{ij}\right)$ and $U\left( x_{i}\right)$ determination.
\section{Acknowledgments}
This work is partially supported by Brazilian agency CNPq (project 151262/2009-8), CAPES (project 
23038002179/2012-50), Universidade Federal do Rio Grande do Sul and Universidade Federal de Santa Catarina.

\bibliography{biblioteca2}
\end{document}